\documentclass[showpacs,pra,preprint,amsmath,amssymb,aps]{revtex4-1}
\usepackage[colorlinks=true,pdfstartview=FitV,linkcolor=red,citecolor=blue,urlcolor=blue]{hyperref}
\usepackage{bm}
\usepackage{graphicx}
\usepackage{booktabs}
\usepackage{slashed}
\usepackage{hyperref}
\usepackage{amssymb,amsmath,amsfonts}
\usepackage{color}
\usepackage[utf8]{inputenc}
\usepackage[caption=false]{subfig}

\renewcommand{\vec}[1]{{\bm #1}}

\newcommand{\acl}{\overline{A}}
\newcommand{\aquf}{\mathcal{A}}
\newcommand{\aqu}{\mathcal{A}_0}

\newcommand{\be}{\begin{equation}}
\newcommand{\ee}{\end{equation}}
\newcommand{\ba}{\begin{eqnarray}}
\newcommand{\ea}{\end{eqnarray}}
\newcommand{\bi}{\begin{itemize}}
\newcommand{\ei}{\end{itemize}}
\newcommand{\la}{\label}

 \newcommand{\eela}[1]{\quad\hbox{\scriptsize{#1}}\label{#1}\end{eqnarray}}
 \newcommand{\eelb}[1]{\label{#1}\end{eqnarray}}

        \def\be{\begin{eqnarray}}    \def\ee{\end{eqnarray}}
 \def\bi#1{\begin{itemize}\item[#1]}     \def\ei{\end{itemize}}  
   \def\^#1{\hat{#1}}

 \def\d{\delta}

 \def\ffract#1#2{\raise .2 em\hbox{$\scriptstyle#1$}\kern-.3em/
                 \kern-.2em\lower .15 em \hbox{$\scriptstyle#2$}}

\def\bmatrix{\begin{matrix}} \def\ematrix{\end{matrix}} \def\bpmatrix{\begin{pmatrix}}\def\epmatrix{\end{pmatrix}}
\def\bcenter{\begin{center}} \def\ecenter{\end{center}}

\def\lowerheightfig#1#2#3{\(\raise-#1\hbox{\includegraphics[height=#2]{#3}}\)}
\def\lowerwidthfig#1#2#3{\(\raise-#1\hbox{\includegraphics[width=#2]{#3}}\)}

\usepackage{color}

\begin{document}

\title{Free energy of a Holonomous Plasma}

\author{Chris P. Korthals Altes}
\affiliation{Aix Marseille Univ, Université de Toulon, CNRS, CPT, Marseille, France}
\affiliation{NIKHEF theory group, P.O. Box 41882, 1009 DB Amsterdam, The Netherlands}

\author{Hiromichi Nishimura}
\affiliation{Research and Education Center for Natural Sciences, Keio University, Toyko, Japan}
\affiliation{RIKEN BNL Research Center, Brookhaven National Laboratory, Upton, NY 11973, USA}

\author{Robert D. Pisarski}
\affiliation{Department of Physics, Brookhaven National Laboratory, Upton, NY 11973, USA}

\author{Vladimir V. Skokov}
\affiliation{Department of Physics, North Carolina State University, Raleigh, North Carolina 27695, USA}
\affiliation{RIKEN BNL Research Center, Brookhaven National Laboratory, Upton, NY 11973, USA}
  
\begin{abstract}
  At a nonzero temperature $T$, a constant field $\acl_0 \sim T/g$ generates nontrivial
  eigenvalues of the thermal Wilson line.
  We discuss contributions to the free energy of such a holonomous plasma when the coupling constant, $g$, is weak.
  We review the computation to $\sim g^2$ by several alternate methods, and show that gauge invariant
  sources, which are nonlinear in the gauge potential $A_0$, generate 
  novel contributions to the gluon self energy at $\sim g^2$.  These ensure the gluon self energy remains
  transverse to $\sim g^2$, and are
  essential in computing contributions to the free energy at $\sim g^3$ for small holonomy, $\acl_0 \sim T$.
  We show that the contribution $\sim g^3$ from off-diagonal gluons is discontinuous
  as the holonomy vanishes. The contribution
  from diagonal gluons is continuous as the holonomy vanishes, but
  sharply constrains the possible sources which generate nonzero holonomy, and must involve
  an infinite number of Polyakov loops.
\end{abstract}

\maketitle

The collisions of heavy nuclei at very high energies demonstrate the existence of a qualitatively new state of
matter.  It is most natural to assume that this is the production of a Quark-Gluon Plasma (QGP) which is, at least
approximately, in thermal equilibrium at a temperature $T$.  The properties of the QGP can be computed
perturbatively in the coupling constant $g$
\cite{Collins:1974ky,*Shuryak:1977ut,Kapusta:1979fh,Toimela:1982hv,Arnold:1994ps,*Zhai:1995ac},
but this is only useful at very high temperature.  At lower temperature, resummation is imperative
\cite{Kajantie:2002wa,Haque:2012my,*Haque:2013sja,*Andersen:2015eoa}, but this again fails
at temperatures several times the transition temperature, which can be termed a ``semi''-QGP.
Numerical simulations on the lattice \cite{Philipsen:2019rjq} provide
detailed information at all temperatures in equilibrium, but at present this is much harder near equilibrium, such
as to compute transport coefficients.

In the pure gauge theory the order parameter for deconfinement are
Polyakov loops.  In an $SU(N)$ gauge theory, up to global $Z(N)$ rotations these are near unity at high
temperature, and, if charged under $Z(N)$, vanish in the confined phase.  Thus the semi-QGP is characterized by
nonzero holonomy for Polyakov loops, where they are nonzero but less than unity.

To treat such a holonomous plasma, it is most natural to take a constant, background field for the vector
potential, $\acl_0 \sim \Theta \, T/g$, where $\Theta$ is a diagonal, traceless color matrix
\cite{Gross:1980br,Weiss:1980rj,Elze:1989rh,Enqvist:1990mm,Belyaev:1991gh,bhattacharya_interface_1991,*bhattacharya_zn_1992,KorthalsAltes:1993ca,Giovannangeli:2002uv,*Giovannangeli:2004sg,altes2007,Dumitru:2013xna,Guo:2014zra,Guo:2018scp}.
In this paper we consider the analysis of a holonomous plasma in perturbation theory.

The computation of the holonomous potential at leading order is reviewed in Sec. (\ref{sec:oneloop}),
mainly to establish notation \cite{Gross:1980br,Weiss:1980rj}.  It is atypical, as
a potential for holonomy first arises then.  The computation at
$\sim g^2$ is given in Sec. (\ref{sec:twoloop})
\cite{Elze:1989rh,Enqvist:1990mm,Belyaev:1991gh,bhattacharya_interface_1991,*bhattacharya_zn_1992,KorthalsAltes:1993ca,altes_potential_2000,Giovannangeli:2002uv,*Giovannangeli:2004sg,altes2007,Dumitru:2013xna,Guo:2014zra,Guo:2018scp,KorthalsAltes:2019yih,altes_nishimura}.
We use several different methods, and show that the potential is only gauge invariant in the presence
of gauge invariant sources involving the Polyakov loops.  Because these are nonlinear functions of the
gauge field, these generate new contributions to the gluon self energy $\sim g^2$.  These are nonlocal,
but essential in showing that the gluon self energy remains transverse to this order.

If the holonomy is large, $\Theta \sim 1$, then the contribution of the off-diagonal gluons
to the free energy is a power series in $g^2$.  If the holonomy is weak,
however, $\Theta \sim g$, then there are contributions to the free energy $\sim g^3$, as in the perturbative vacuum
\cite{Kapusta:1979fh}.  Previously we demonstrated that a novel result occurs at this order
\cite{KorthalsAltes:2019yih}: the contribution from off-diagonal gluons jumps discontinuously as
the holonomy goes to zero. In Sec. (\ref{sec:cubic}) we demonstrate this surprising result by another
more direct means from that in Ref. \cite{KorthalsAltes:2019yih}, using Hard Thermal Loops \cite{Hidaka:2009hs}.

In Sec. (\ref{sec:diagonal}) we show
that while the contribution from diagonal gluons vanishes smoothly with the holonomy,
that this requires rather nontrivial constraints on the associated sources.
We are able to establish rigorous constraints for two \cite{Dumitru:2012fw, Nishimura:2011md}
and an infinite number of colors
\cite{Dumitru:2004gd,Pisarski:2012bj,Nishimura:2017crr}.  For the latter we use
methods from matrix models
\cite{Brezin:1977sv, *Gross:1980he, *Wadia:1980cp, Lang:1980ws, *Menotti:1981ry, *Jurkiewicz:1982iz,*Green:1983sd, *Damgaard:1986mx, *Azakov:1986pn,*Demeterfi:1990gb,*Jurkiewicz:1990we, *Sundborg:1999ue,*Aharony:2003sx, *Aharony:2005bq,*AlvarezGaume:2005fv,*Schnitzer:2004qt,*Hollowood:2009sy, *Hands:2010zp, *Hollowood:2011ep, *Hollowood:2012nr,*Ogilvie:2012is,*Liu:2015yaa}.
The conclusion is that for the holonomy to turn on smoothly for a weak source, that the source must involve
a sum over an {\it infinite} number of Polyakov loops.

An analysis with the insertion method is treated separately \cite{altes_nishimura}.  This allows one to show that
the free energy is continuous to $\sim g^4$ as the holonomy vanishes.

Understanding the behavior of a holonomous plasma is of intrinsic interest
in understanding the behavior of gauge theories at nonzero temperature.
It is also of use in developing effective theories, which can then be analytically continued to compute
properties near equilibrium
\cite{pisarski_quark_2000,*dumitru_degrees_2002,*dumitru_two-point_2002,*scavenius_k_2002,*dumitru_deconfining_2004,*dumitru_deconfinement_2005,*dumitru_dense_2005,*oswald_beta-functions_2006,*pisarski_effective_2006,*dumitru_eigenvalue_2008,*smith_effective_2013,dumitru_how_2011,*dumitru_effective_2012,sasaki_effective_2012,pisarski_gross-witten-wadia_2012, *lin_zero_2013,kashiwa_critical_2012,*kashiwa_roberge-weiss_2013,*gale_production_2015,*hidaka_dilepton_2015,*satow_chiral_2015,*lin_collisional_2014,Pisarski:2016ixt,*Folkestad:2018psc,KorthalsAltes:2019yih}.
These effective theories involve a perturbative potential for the holonomous potential, in addition
to a non-perturbative term, added by hand, which drives the transition to confinement.  Thus
the present analysis will help in refining such effective theories.  Notably,
the source used as a non-perturbative holonomous potential in these models satisfies that required by the
analysis of Sec. (\ref{sec:diagonal}).

While in this paper we do not consider dynamical quarks, their contribution to the holonomous
potential can be computed directly, including at nonzero density
\cite{altes_potential_2000,Guo:2018scp}.  Doing so, one finds
that the the effective theory developed for the pure gauge theory gives a reasonable analysis of
QCD, with three flavor of light quarks \cite{Pisarski:2016ixt,*Folkestad:2018psc}.  

\section{One loop order}\label{sec:oneloop}

To compute the effective potential one can either use an external source or a constrained path integral.
Of course these must be equivalent, but this is not evident at two loop order
and beyond. 

\subsection{External source}
\label{one_loop_pert_source}

In the presence of an external source $J_\mu$, the Lagrangian for a gauge field is
\begin{equation}
  {\cal L} = \frac{1}{2} \, {\rm tr} \; G^2_{\mu \nu} + {\rm tr} \; J^\mu A_\mu \; , \;
  G_{\mu \nu} = \partial_\mu A_\nu - \partial_\nu A_\mu - i g  [ A_\mu , A_\nu] \; .
\end{equation}
We consider a $SU(N)$ gauge theory, with the generalization to other gauge groups direct.

At a nonzero temperature $T$ gauge invariant quantities are given by
traces of powers of the thermal Wilson line, which are Polyakov loops:
\begin{equation}
  \ell_r(x) = \; \frac{1}{N} \; {\rm tr} \; {\bf L}^r(x) \;\; ; \;\;
  {\bf L}(x) = {\rm tr} \; {\cal P} \; \exp\left(i g \, \int^{1/T}_0 A_0(x,\tau) \, d\tau \right) \, .
\end{equation}
In a holonomous plasma we expand the gauge potential about a classical field, $\acl_0$, and a quantum
field, $\aquf_\mu$,
\begin{equation}
  A_\mu = \acl_\mu + \aquf_\mu \; \; , \; \; \acl_\mu  = \delta_{\mu 0} \; \Theta  \; \frac{T}{g} \; .
\end{equation}
The classical field $\acl_0$ is constant, with
$\Theta$ a diagonal, traceless matrix: $\Theta^{ a b} = \theta^a \delta^{a b}$, $\sum_{a=1}^N \theta^a = 0$.
In this background,
\begin{equation}
  \overline{{\bf L}}^r = {\rm e}^{i r \Theta} \; .
\end{equation}

We use background field gauge, with the gauge dependent terms 
\begin{equation}
  {\cal L}_{\rm gauge}=
  \frac{1}{\xi} {\rm tr} (\bar{D}_\mu \aquf_\mu)^2 + \overline{\eta} \left( - \overline{D}_\mu D_\mu \right) \eta
  \; ,
  \label{gaugefixedaction}
\end{equation}
where $D_\mu = \partial_\mu - i g [A_\mu,]$ and $\overline{D}_\mu = \partial_\mu - i g [\acl_\mu,]$.

The generating functional ${\cal W}(J)$ is defined by
\begin{equation}
  \exp({\cal W}(J)) = \int {\cal D}A_\mu \; {\cal D}\eta  \; {\cal D}\overline{\eta} \;
  \exp\left( - \int^{1/T}_0 d\tau \int d^3 x \; \left( {\cal L} + {\cal L}_{\rm gauge} \right) \right) \; .
\end{equation}
To one loop order the computation proceeds by integrating over the $A_\mu^{\rm qu}$ to quadratic order.  This
is standard, and we only wish to make the following comments.

Assume that the background field is nontrivial, such as for an instanton.  Then the associated field
strength $G_{\mu \nu} \sim 1/g$, and the equation of motion is
\begin{equation}
  \overline{D}_\mu  \overline{G}_{\mu \nu} = J_\nu \; .
\end{equation}
For this to be consistent, $J_\nu \sim 1/g$.

For a constant, diagonal $\acl$ field, though, the classical field strength vanishes identically.
We then assume that the source is not $\sim 1/g$, but $\sim 1$.  As we show later, the source doesn't contribute at leading
order, but it does at next to leading order.

Integrating over $Q_\mu$, we obtain the effective action to one loop order
\begin{equation}
  {\cal S}_{\rm eff} =  - \; {\rm tr} \; \log \left( -(\overline{D}_\mu)^2 \right) \, .
\end{equation}
This is the determinant from gluon and ghost fields.  It can be shown rather directly that this
expression is independent of the gauge fixing parameter, $\xi$.

Define
\begin{equation}
  {\cal W}(J) = \int d^3 x \; (- V(A^{\cal}) + {\rm tr} J_\mu A_\mu ) \; .
\end{equation}
At one loop order, the holonomous potential is
\begin{equation}
  {\cal V}_{1}(\Theta) = \, - \; \frac{T^3}{\pi^2} \; 
  \sum_{n=1}^\infty \frac{1}{n^4} \; \left|{\rm tr} \, \overline{{\bf L}}^{\, n} \right|^2 = 
  \frac{2 \pi^2 T^3}{3} \; \sum_{a,b = 1}^N B_4\left(\frac{\theta_a-\theta_b}{2 \pi} \right) \; ,
  \label{potential_one_loop}
\end{equation}
where $B_4$ is the fourth Bernoulli polynomial,
\begin{equation}
  B_4(x) = - \, \frac{1}{30} + x^2 (1 - |x|)^2 \; .
  \label{def_B4}
\end{equation}
In Eq. (\ref{potential_one_loop}) $\theta_a - \theta_b$ is defined modulo $2\pi$, since in the
thermal Wilson line the $\theta_a$ are angular variables.

The one particle irreducible (1PI) generating functional is the Legendre transformation of ${\cal W}(J)$,
\begin{equation}
  \Gamma(\widetilde{A}_0)
  = {\it sup} \left( \int d^3 x \; {\rm tr} ( J_\mu \widetilde{A}_0 ) - {\cal W}(J) \right) \; .
\end{equation}
Here ${\it sup}$ denotes that one finds the point extremal with respect to variations in $J$.
Usually the field is a function of $J$.  
Here because of the degeneracy at leading order, though, $\widetilde{A}_0$ is independent of $J$.
Hence the variation 
 is trivial, and simply 
imposes $\widetilde{A}_0 = \acl_0$, giving
\begin{equation}
  \Gamma(\widetilde{A}_0) = {\cal V}_1(\Theta) \; .
  \label{zero_loop}
\end{equation}
Note the distinction with the usual effective potential: we do not use the equations of motion to
require that the variation of ${\cal V}_1(\acl_0(J))$ is extremal with respect to $J$.

Beyond leading order, the potential ${\cal V}_1(\Theta)$ lifts the degeneracy. 

\subsection{Constrained functional integral}\label{sec:constraint}

Another way of computing is to constrain the value of the spatial average of the Polyakov loop.
To avoid clutter we constrain only $\ell_1$, with the complete generalization given below, Eq. (\ref{general_constraint}).
The constrained functional integral is
\begin{equation}
  \exp(- \, V \, {\cal V}(\overline{\ell})) = \int {\cal D}A_\mu \; {\cal D}\eta  \; {\cal D}\overline{\eta} \;
  \delta\left( \overline{\ell} - \int \frac{d^3 x}{V} \ell_1(x) \right ) 
  \exp\left( - \int^{1/T}_0 d\tau \int d^3 x \; \left( {\cal L} + {\cal L}_{\rm gauge} \right) \right) \; .
\la{pathintwithconstraint}
\end{equation}
We exponentiate the constraint,
\begin{equation}
  \delta\left( \overline{\ell} - \int \frac{d^3 x}{V} \ell_1(x) \right )
  = \int d \epsilon  \; \exp\left( i \epsilon \left(\overline{\ell} - \int \frac{d^3 x}{V} \ell_1(x)\right) \right) \; .
\end{equation}
Since we constrain only the spatial average of the loop(s), there is only a single constraint field, $\epsilon$.
$V$ is the spatial volume.

We expand the constraint field,
\begin{equation}
  \epsilon = \epsilon^{\rm cl} + \epsilon^{\rm qu} \; .
  \label{single_epsilon}
\end{equation}
A nonzero value of $\epsilon^{\rm cl}$ acts like an external source.  Since there is no potential for
$\acl$ at leading order, this source vanishes at leading order,
\begin{equation}
  \epsilon^{\rm cl} = 0 \; .
  \label{insertion}
\end{equation}

As before, it is direct to compute the constrained partition function.  Integrating over $\epsilon^{\rm qu}$
imposes the constraint on the loop, requiring $\acl$ to give the requisite value of the loop, $\ell_1$.
Integration over $A_\mu^{\rm qu}$ is trivial, because the equation of motion vanishes anyway.  The integration
over $A_\mu^{\rm qu}$ is also unaffected, as the constraint field doesn't contribute, $\epsilon^{\rm cl} = 0$.
The result is that of Eq. (\ref{zero_loop}).

\section{Two loop order}\label{sec:twoloop}

The computation of the free energy to $\sim g^2$ is an old story~
\cite{Elze:1989rh,Enqvist:1990mm,Belyaev:1991gh,bhattacharya_interface_1991,*bhattacharya_zn_1992,KorthalsAltes:1993ca,Giovannangeli:2002uv,*Giovannangeli:2004sg,altes2007,Dumitru:2013xna,Guo:2014zra,Guo:2018scp}.  Nevertheless, as we show
there are subtleties in the computation of the gluon self energy to the same order.  Thus we summarize the
computation briefly in order to introduce behavior of the gluon self energy to this order.  This is
essential in order to compute corrections to higher loop order, starting at $\sim g^3$.

\subsection{Linear gauge}

To two loop order, the result for the potential is
\begin{equation}
  {\cal V}^{\rm pert}_2(\Theta) = \frac{g^2\, T^3}{4} \; \sum_{a,b,c=1}^N \;
  B_2\left(\frac{\theta_a - \theta_c}{2 \pi}\right) \;
  B_2\left(\frac{\theta_b - \theta_c}{2 \pi}\right) + (1 - \xi)
  B_1\left(\frac{\theta_a - \theta_c}{2 \pi}\right) \;
  B_3\left(\frac{\theta_b - \theta_c}{2 \pi}\right) \; ,
  \label{two_loop_pert_pot}
\end{equation}
This involves the first, second, and third Bernoulli polynomials,
\begin{align}
  B_1(x) &= - \frac{1}{2} \; {\rm sign}(x) + x \;\; ; \nonumber \\
  B_2(x) &= \frac{1}{6} - |x| + x^2 \;\; ; \nonumber \\
  B_3(x) &= \frac{1}{2} \; x - \frac{3}{2} \; {\rm sign}(x) \; x^2 + x^3 \; .
  \label{other_bernoullis}
\end{align}
Each difference of the $\theta$'s, such as $\theta_a - \theta_b$, is
defined modulo $2\pi$.  Even Bernoulli polynomials are even in $x$, and so depend
only upon $|\theta_a - \theta_b|$.  Odd Bernoulli polynomials are odd in $x$.

The potential in Eq. (\ref{two_loop_pert_pot})
is rather unexpected since it explicitly depends on $\xi$.  It can also be
shown that there is a minimum at a nonzero value of $q \sim (3 - \xi) g^2$. 
Note however that the pressure is $\xi$ independent to torder $g^2$. 	 

The $\xi$ dependence can be understood from the Nielsen identities
\cite{Nielsen:1975fs}.  For a value of
$\theta \sim g^2$, it contributes to the potential at $\sim g^4$.
Nevertheless, it is useful to see how this a gauge invariant result arises explicitly.  Doing so
we show that the usual perturbative vacuum is stable.

\subsection{Constrained functional integral}
\label{sec:constraintg2}

Since the above source and potential are gauge variant, we introduce gauge invariant constraints into the action.
For $SU(N)$ we constrain the Polyakov loops by adding constraint fields $\epsilon_r$ to the action,
\begin{equation}
  {\cal S}_{\rm cons} = i \sum_{r = 1}^{N} \epsilon_r \left(
    \ell_r - \int \frac{d^3 x}{V} \; {\rm tr} \; {\bf L}^r(x) \right) \; .
  \label{general_constraint}
\end{equation}
Only $N-1$ constraints are needed, but we find it convenient to use one too many constraints, from
$r = 1$ to $N$ instead of $N-1$.  This is done for the following reason.  For $SU(N)$ the
sum of the $\theta_a$'s vanishes, and there are only $N-1$ independent $\theta_a$'s.  It is awkward to eliminate
one of the $N$ $\theta_a$'s in favor of the independent variables, though.  Instead, it is easier to
pretend as if all of the $N$ $\theta_a$ are independent, and derive the equations of motion for the $N$ $\theta_a$.

\subsubsection{Insertion method}
\label{sec:insertion}

The insertion method is a straightforward expansion of the gauge action and the constraint in terms of the
fluctuation fields ${\cal A}_\mu$ and $\epsilon^{\rm qu}$
\cite{KorthalsAltes:1993ca,Giovannangeli:2002uv,*Giovannangeli:2004sg,altes2007,Dumitru:2013xna,Guo:2014zra,Guo:2018scp,altes_nishimura}.
This gives constant terms, linear terms, quadratic terms, and interaction terms.
The linear terms are set to zero, and fixes $\epsilon^{\rm cl} = 0$, as in 
Eq. (\ref{insertion}), and $\bar A$ in terms of $\ell$.
The quadratic terms in the action now include
$
{\cal L}_{quadr}-i\epsilon^{\rm qu}{\cal A}_0(0)
$,
where ${\cal A}_0(0)$ is the zero momentum component of the fluctuation field.
Vertices are generated by expanding the gauge action plus the constraint, 
\begin{equation}
{\cal L}_{int} +i\epsilon^{\rm qu}\left({\bf L}_2+{\bf L}_3+... \right) \; ,
\end{equation}
where the subscripts indicate the powers of the quantum fluctuation ${\cal A}_\mu$.  
Then the integration over $\epsilon^{\rm qu}$ is done. This reinstates the delta function
of the original constraint but now in the simple form $\delta({\cal A}_0(0))$ times
the pure gauge field vertices. It also introduces 
also new vertices, where $\epsilon^{\rm qu}$ multiplies
${\cal L}_{int}\left({\bf L}_2+...\right) $;
this generates derivatives of the delta-function. 
The derivatives in ${\cal A}_0(0)$ act through integration by parts on the gauge interaction vertices
and on the Polyakov loops. These are called the insertion vertices:
integration over the remaining fluctuations gives then, apart from the usual 
$QCD$ diagrams, "insertion diagrams"
\cite{KorthalsAltes:1993ca,Giovannangeli:2002uv,*Giovannangeli:2004sg,altes2007,Dumitru:2013xna,Guo:2014zra,Guo:2018scp,altes_nishimura}.  These are key to understanding how gauge invariance is implemented.  The insertion terms do generate
contribution to the two, three, and higher point functions of the gluons.
Up and including three loop order
the thermodynamic limit  poses no problems, except in the case
of  diagonal gluons with two self energy insertions. There, the finite size
corrections to the self energy  have to be taken into account.

\subsubsection{Alternate approach}
\label{sec:alternate}

In the insertion approach, $\epsilon^{\rm cl} = 0$, Eq. (\ref{insertion}), order by order in perturbation
theory.  An alternate approach is the following.
Since the degeneracy in $\Theta$ is broken at one loop order, we generalize Eq. (\ref{potential_one_loop})
from a function of the background field, $\acl_0$, to a function of the full vector potential,
$A_\mu = \acl_\mu + \aquf_\mu$:
\begin{equation}
  {\cal V}_{1}(A_0) = \, - \; \frac{T^3}{\pi^2} \; \int d^3 x \;
  \sum_{n=1}^\infty \frac{1}{n^4} \; \left|{\rm tr} \, {\bf L}^{\, n}(x) \right|^2  \; .
  \label{potential_one_loop_full}
\end{equation}
We then add and subtract ${\cal V}_{1}(A_0)$ to the Lagrangian.  The subtracted term cancels ${\cal V}_{1}(\acl_0)$
the same term when it is generated at one loop order.  This is exactly analogous to how, for example, a Debye mass is included
in perturbation theory.

The advantage of adding ${\cal V}_1(A_0)$ is that the degeneracy with respect
to $\acl_0$, valid at the classical level, is lifted.  The equations of motion are now
\begin{equation}
  - \; \frac{i }{V} \sum_{r=1}^N  \;  i \, \epsilon^{\rm cl}_r \;
  r \;  {\rm e}^{ i r \theta_a}
  = \frac{8 \pi T^3}{3} \sum_{b=1}^N B_3\left(\frac{\theta_a - \theta_b}{2 \pi}\right) \; ;
  \label{constraint_eom}
\end{equation}
$B_3$ is the third Bernoulli polynomial, which arises as the derivative of $B_4(x)$.
There are $N$ equations of motion in Eq. (\ref{constraint_eom}).  As an odd Bernoulli polynomial, $B_3(x)$ is defined
to be odd in $x$, and so by summing over $a$, we obtain
\begin{equation}
  \sum_{r=1}^N \sum_{a=1}^N r \; \epsilon^{\rm cl}_r \; {\rm e}^{i r \theta_a} = 0 \; .
  \label{equation_motion_constraint}
\end{equation}
In principle it is possible to eliminate one of the $\theta_a$'s for the $N-1$ independent variables.
As we shall see, however, unexpectedly there is no need to explicitly do so, nor to solve for
the values of the constraint fields $\epsilon_r$.  This greatly simplifies matters.

The equation of motion in Eq. (\ref{constraint_eom}) is identical to that with an external source
${\cal J}_r$ which couples to the Polyakov loop $\ell_r$, with $\epsilon_r^{\rm cl} = i {\cal J}_r \, V$.
This was the approach used in our previous work \cite{KorthalsAltes:2019yih}.
With a constraint action, it is natural that the expectation value of the classical field is
imaginary and proportional to the spatial volume.   Also notice that the source ${\cal J}_r$ are naturally of order
one, and not $\sim 1/g$, in agreement with the analysis in Sec. (\ref{one_loop_pert_source}).
With either a constraint or a source, however, it is necessary to explicitly add the one loop
term to life the degeneracy in $\Theta$.  Thus adding 
Eq. (\ref{potential_one_loop_full}) above is equivalent to Eq. (18) of Ref. \cite{KorthalsAltes:2019yih}.

\subsection{Expansion of Polyakov loops to quadratic order}\label{sec:polloopg2}

The major difference between a source that couples to Polyakov loops, and the usual term which is
linear in $A_\mu$, is that Polyakov loops are an infinite power series in $\aqu$.  
To $\sim g^2$, it is necessary to include terms of quadratic order in
$\aqu$, and so on to higher order.  In this subsection we compute the terms to quadratic
order.  

We need the thermal Wilson for a time of limited extent, $\tau': 0 \rightarrow \tau$, 
\begin{equation}
  {\bf L}(x,\tau) = {\cal P} \exp\left( i g \int^\tau_0 A_0(x,\tau') d\tau'\right) \; .
\end{equation}
where ${\cal P}$ denotes path ordering.  For the $r^{th}$ power of the Wilson line,
\begin{equation}
  {\rm tr} \, {\bf L}^r(x,1/T) = {\rm tr} \, {\bf L}(x,r/T) \; .
\end{equation}

We define the expansion about the classical field as
\begin{equation}
  {\bf L}^r(x,1/T) = \overline{{\bf L}}^r + \delta {\bf L}_1^r(x) + \delta {\bf L}_2^r(x) + \ldots \;\; ; \;\;
  \overline{{\bf L}}^r = {\rm e}^{i \, r \, \Theta} \; ,
\end{equation}
where the subscript denotes the power of $\aqu$.

To linear order,
\begin{equation}
  \delta {\bf L}^r_1(x) = i g \int^{r/T}_0 d\tau \;
  \overline{{\bf L}}(r/T - \tau) \aqu(x,\tau) \overline{{\bf L}}(\tau) \; .
\end{equation}
Taking the trace,
\begin{equation}
  {\rm tr} \; \delta {\bf L}_1^r(x)
  = i g \, r \; {\rm tr} \left( {\rm e}^{ i r  \Theta} \int^{1/T}_0 d \tau \, \aqu(x,\tau) \right) \; .
  \label{first_exp_loop}
\end{equation}
As $\overline{{\bf L}}$ is diagonal, only diagonal elements of $\aqu$
contribute.  The integral over $\tau$ projects out the constant mode in $\tau$ for $\aqu(x,\tau)$.
To derive the equations of motion, it is useful to shift $\theta_a \rightarrow \theta_a + \delta \theta_a(x)$, so that
\begin{equation}
  {\rm tr} \; \delta {\bf L}_1^r(x)
  = \sum_{a=1}^N  i \, r \; {\rm e}^{ i r  \theta_a} \; \delta \theta_a(x) \; ,
\label{firstorderL}
\end{equation}
which gives the left hand side of Eq. (\ref{constraint_eom}).

To proceed further we need to choose an explicit basis.  We adopt the double line notation familiar at large $N$
to finite $N$.  In the fundamental representation, 
\begin{equation}
  \left(t^{a b}\right)_{c d}
  = \frac{1}{\sqrt{2}} \left( \delta^{a c} \delta^{b d} - \frac{1}{N} \delta^{a b} \delta^{c d} \right) \; ,
\end{equation}
$a,b,c,d\ldots = 1 \ldots N$.  An adjoint
matrix is denoted by the pair of upper indices, $a b$.  Hence there is one too many generators,
$N^2$ in all instead of $N^2 - 1$.  The normalization of off-diagonal generators is standard, 
\begin{equation}
  {\rm tr} (t^{a b} t^{b a}) = \frac{1}{2} \;\; , \; \; a \neq b \; .
  \label{norm_off_diag}
\end{equation}
Because the double lines of $SU(N)$ are ordered in opposite directions, the indices flip
when two generators are contracted.

This basis is overcomplete by one diagonal generator.  Consequently
the normalization of the diagonal generators is unusual, Eqs. (16) and (17) of Ref. \cite{Hidaka:2009hs}.
However, it is easy just multiplying diagonal matrices together, and so at least to the order
at which we work, this can be ignored.

For example, to quadratic order the diagonal elements are
\begin{equation}
   (\delta{\bf L}_2^r)_{\rm diag} = - g^2 
\int^{r/T}_0 d\tau_1 \int^{\tau_1}_0 d \tau_2 \; 
{\rm e}^{i r \theta_a} \, \aqu^{aa}(x,\tau_1)   \, \aqu^{aa}(x,\tau_2) \; .
\label{quadratic_diagonal}
\end{equation}
For the modes constant in time path ordering doesn't matter, and this is elementary.  Path ordering
does enter for time dependent modes.

More interesting are the off-diagonal elements:
\begin{equation}
  \left( \delta {\bf L}_2^r\right)_{\rm off} = - g^2 \int^{r/T}_0 d\tau_1 \int^{\tau_1}_0 d \tau_2 \; 
  \overline{{\bf L}}(r/T - \tau_1) \, \aqu(x,\tau_1) \, \overline{{\bf L}}(\tau_1 - \tau_2)
  \, \aqu(x,\tau_2) \, \overline{{\bf L}}(\tau_2) \; .
\end{equation}
For each of the $\aqu$'s we go from the imaginary time $\tau$ to momentum space,
$$ \aqu(x,\tau_1) = T \sum_{n= -\infty}^{+\infty}
{\rm e}^{-i p_0 \tau_1} \aqu(x,p_0) \; , \; p_0 = 2 \pi n T $$
\begin{equation}
  \aqu(x,\tau_2) = T \sum_{n'= -\infty}^{+\infty}
  {\rm e}^{-i p_0' \tau_2} \aqu(x,p_0') \; , \; p_0' = 2 \pi n' T \;.
\end{equation}
Because the Wilson line is nonlocal in time, it is 
possible that terms where $p_0 \neq p_0'$ contribute.
In contrast, since the terms are local in space, the spatial momenta
of the two $\aqu$'s are equal and opposite.

The color structure enters in two ways:
\begin{equation}
  \aqu^{ba}(p_0) t^{ab} \; \aqu^{ab}(p_0') t^{ba} \;\; ; \;\;
  \aqu^{ ab}(p_0) t^{ba} \; \aqu^{ba}(p_0') t^{ab} \; .
  \label{two_terms}
\end{equation}
There is no summation over repeated indices, as the color indices $a$ and $b$, with $a \neq b$, are fixed.

Begin with the first permutation.  Since $\overline{{\bf L}}$ is a diagonal matrix,
\begin{equation}
  \overline{{\bf L}}(\tau) \; t^{a b} = {\rm e}^{i \theta_a \tau T} \; t^{a b} \;\; ; \;\;
    t^{a b} \; \overline{{\bf L}}(\tau)   =  t^{a b} \; {\rm e}^{ i \theta_b \tau T}  \; .
\end{equation}
Thus the first permutation in Eq. (\ref{two_terms}) gives
\begin{equation}
  - g^2 T^2 {\rm e}^{ i r \theta_a} \sum_{n,n'=-\infty}^{+\infty} \int^{r/T}_0 d\tau_1
  \; {\rm e}^{-i p_0^{a b} \tau_1} \;
  \int^{\tau_1}_0 d \tau_2 \;  {\rm e}^{-i p_0'^{b a} \tau_2} \;
  \left( \aqu^{ba}(p_0) t^{ab} \; \aqu^{ab}(p_0') t^{ba} \right) \; ,
\end{equation}
where
\begin{equation}
  p_0^{a b} = T (2 \pi n + \theta_a - \theta_b) \; , \;
  p_0'^{b a} = T (2 \pi n' + \theta_b - \theta_a) \; .
  \label{define_p0}
\end{equation}
The integral over $\tau_2$ is 
\begin{equation}
  \int^{\tau_1}_0 d \tau_2 \;  {\rm e}^{-i p_0'^{b a} \tau_2} =
  \frac{1}{- i p_0'^{b a}} \left( {\rm e}^{-i p_0'^{b a} \tau_1} - 1 \right) \; .
\end{equation}
Integrating over $\tau_1$,
\begin{equation}
  - g^2 T^2 \sum_{n,n'=-\infty}^{+\infty} \left( r \, {\rm e}^{2 \pi i r q_a} \, \frac{1}{- i p_0'^{b a}} \delta(p_0 + p_0')
    + \frac{1}{p_0^{a b}p_0'^{b a}} \left( {\rm e}^{i r \theta^b} - {\rm e}^{ i r \theta^a} \right) \right) \;
  \left( \aqu^{ba}(p_0) t^{ab} \; \aqu^{ab}(p_0') t^{ba} \right) \; .
   \label{quadratic_wilson_lineA}
\end{equation}

The other ordering in Eq. (\ref{two_terms}) gives
\begin{equation}
  - g^2 T^2 \sum_{n,n'=-\infty}^{+\infty} \left( r \, {\rm e}^{ i r \theta_b}\, \frac{1}{i p_0^{a b}} \delta(p_0 + p_0')
    + \frac{1}{p_0^{a b}p_0'^{b a}} \left( {\rm e}^{ i r \theta^a} - {\rm e}^{ i r \theta^b} \right) \right) \;
  \left( \aqu^{ ab}(p_0') t^{ba} \; \aqu^{ ba}(p_0) t^{ab} \right) \; ,
  \label{quadratic_wilson_lineB}
   \end{equation}
where we relabel $p_0 \leftrightarrow p_0'$ and $a \leftrightarrow b$.
This agrees with previous results, such as Eq. (3.12) of \cite{KorthalsAltes:1993ca}.

We now add the two orderings. 
With the normalization of Eq. (\ref{norm_off_diag}), we find for the sum of off-diagonal elements
\begin{equation}
  {\rm tr}\, \left( \d{\bf L}_2^r(x)\right)_{\rm off} =  - \, \frac{g^2}{4 i } \sum_{a \neq b = 1}^N
r \, \left( {\rm e}^{i r \theta_a} - {\rm e}^{ i r \theta_b} \right) \, 
T \sum_{n=-\infty}^{+\infty}\,   \frac{1}{p_0^{a b}}
  \left( \aqu^{ba}(x,p_0) \; \aqu^{ ab}(x,-p_0) \right) \; .
  \label{quadratic_r}
\end{equation}
The second terms in Eqs. (\ref{quadratic_wilson_lineA}) and (\ref{quadratic_wilson_lineB})
are truly nonlocal in time, as $p_0 + p'_0 \neq 0$ contribute.  After taking the trace, however,
these terms cancel: because of the energy denominator, the result is diagonal in $p_0$ and non-local in
the Euclidean time.

This term is special to the off-diagonal modes.  For example, for the diagonal modes which
are constant in time, $p_0 = p_0' = 0$, Eq. (\ref{quadratic_diagonal}) reduces to
\begin{equation}
  - g^2 {\rm e}^{ i r \theta_a} r^2 \left(\aqu^{aa}\right)^2 \; .
\end{equation}

\subsection{Corrections to Polyakov loops and the free energy at $\sim g^2$}\label{sec:2loopeffpot}

The result in Eq. (\ref{quadratic_r}) is useful in several ways.
We first show how the results above can be used to compute the free energy
at $\sim g^2$ in two different, but equivalent ways.  This is necessary
to compute corrections at higher order, to $\sim g^3$.

Consider the constraint action of Eq. (\ref{general_constraint}).  As in Eq. (\ref{single_epsilon}),
we expand the $N$ constraint fields in classical and quantum components,
\begin{equation}
  \epsilon_r = \epsilon^{\rm cl}_r + \epsilon^{\rm qu}_r \; .
  \label{general_epsilon}
\end{equation}
For each of the $r$ constraint fields, the classical value of $\epsilon_r^{\rm cl}$ is determined
by varying with respect to $\aqu$, and is given by Eq. (\ref{constraint_eom}).

The terms linear in $\epsilon_r^{\rm qu}$ are
\begin{equation}
  i \, \sum_{r=1}^N \epsilon_r^{\rm qu}  \left( \ell_r - \int \frac{d^3 x}{V} \; {\rm tr} \left( \overline{{\bf L}}^r
      + \delta {\bf L}_{2}^r(x) + \ldots \right) \right) - i \, \sum_{r=1}^N \epsilon_r^{\rm qu} \; 
  \int \frac{d^3 x}{V} \; {\rm tr} \; \delta {\bf L}_{1}^r(x)    \; .
  \label{expand_constraint}
\end{equation}
Consider the last term, which is quadratic in the quantum fields, $\sim \epsilon_r^{\rm qu} \aqu$.
From the form of $\delta {\bf L}_1^r$ in Eq. (\ref{first_exp_loop}), only the static, $p_0 = 0$ component of
$\aqu$ enters.  Further, the constraint is over the spatial average of $\delta {\bf L}_1^r$, which
further projects out the zero momentum component of the spatial momentum, $\vec{p} = 0$.  Unlike the static
component in $p_0$, which is of finite measure, the zero momentum component in $\vec{p}$ is of zero measure.
Thus we can ignore this part of the integral over $\epsilon_r^{\rm qu}$.

To evaluate Eq. (\ref{expand_constraint}) we use Eq. (\ref{quadratic_r}) to find
\begin{equation}
  \langle {\rm tr}\, \left(\d{\bf L}_2^r(x)\right) \rangle_{\rm off} = (3 - \xi) \; \frac{g^2}{8 \pi} \; \sum_{a \neq b = 1}^N
  i \, r \, {\rm e}^{ i r \theta_a} \; B_1\left(\frac{\theta_a - \theta_b}{2 \pi}\right) \; .
  \label{corrected_second_loop}
\end{equation}
This first term in Eq. (\ref{expand_constraint}) has a simple physical interpretation:
as discussed by Belyaev \cite{Belyaev:1991gh}, it represents a correction to the constraint
at $\sim g^2$.  This can be implemented by going from a ``bare'' $\theta_a$ to a renormalized $\theta_a$.
This shift is finite, but field and $\xi$ dependent.  Using this shifted $\theta_a$ in the
free energy obtained perturbatively, one obtains the result below, Eq. (\ref{final_two_loop_pot}).

The shift in the $\theta_a$'s is natural.  While the thermal Wilson line is gauge dependent, 
the eigenvalues of the Wilson line are gauge invariant.  Shifting the eigenvalues is
one way of implementing this.

The same result is obtained by using the insertion method of Sec. (\ref{sec:insertion}).  There
instead of a shift in the eigenvalues, there are new diagrams from expanding the constraint
\cite{KorthalsAltes:1993ca,Giovannangeli:2002uv,*Giovannangeli:2004sg,altes2007,Dumitru:2013xna,Guo:2014zra,Guo:2018scp,altes_nishimura}.

Lastly, there is a third method of computing the free energy to $\sim g^2$.
With the method of Sec. (\ref{sec:alternate}), to $\sim 1$ the
constraint field develops an expectation value, $\epsilon^{\rm cl}_r \neq 0$.  
Using Eq. (\ref{constraint_eom}), the quadratic term in Eq. (\ref{quadratic_r}) contributes to the action as
\begin{equation}
  - \, \frac{4 \pi }{3} \, g^2 T^3 \, \sum_{a,b,c=1}^N  \left( B_3\left(\frac{\theta_a - \theta_c}{2 \pi}\right)
    - B_3\left(\frac{\theta_b - \theta_c}{2 \pi}\right) \right)
  \;  T \sum_{n=-\infty}^{+\infty}  \frac{1}{p_0^{a b}} \int d^3 x  
     \left( \aqu^{ ba}(x,p_0) \; \aqu^{ ab}(x,-p_0) \right) \; .\\
\label{quadratic_redux2}
\end{equation}
Notice that the factor of $1/V$ in the constraint is compensated by $\epsilon_r^{\rm cl} \sim V$ in
Eq. (\ref{constraint_eom}).
Rather surprisingly, this constribution is completely independent of the detailed
form of the $\epsilon_r^{\rm cl}$: once the equations of motion are imposed,
they completely drop out.  
We generalize this later to sources involving two traces in Sec. (\ref{double_trace_constraint}).

By contracting two $\aqu$ fields together,  Eq. (\ref{quadratic_redux2}) contributes to
the holonomous potential.
The result is gauge variant, and proportional to $\xi$,
\begin{equation}
  {\cal V}_2^{\rm cons}(\Theta) = - (3 - \xi) \; \frac{g^2 T^3}{3}\; \sum_{a,b,c=1}^N
  B_1\left(\frac{\theta_a - \theta_c}{2 \pi}\right) \; B_3\left(\frac{\theta_b - \theta_c}{2 \pi}\right) \; .
  \label{cons_two_loop_pot}
\end{equation}
After some juggling \cite{Dumitru:2013xna, Guo:2014zra} of Bernoulli polynomials, 
\begin{equation}
  {\cal V}_2(\Theta) = {\cal V}_2^{\rm pert}(\Theta) + {\cal V}_2^{\rm cons}(\Theta) = - \frac{5 }{24}  \; g^2 T^3 \;
  \sum_{a,b = 1}^N B_4\left(\frac{\theta_a - \theta_b}{2 \pi}\right) \; .
  \label{final_two_loop_pot}
\end{equation}
This is both independent of the gauge fixing parameter, $\xi$, and proportional to the potential
at one loop order.  As such, the perturbative vacuum $\Theta = 0$ is stable.

Each $B_n$ can be written as a sum of double traces of the Wilson line.  Thus the terms
$\sim B_2 B_2$ and $\sim B_1 B_3$ involve four traces.  The final form 
$\sim B_4$, though, only involves
two traces.  This has interesting implications for the solutions of the theory at infinite $N$,
where potentials with only double traces are often soluble, at least in certain limits.

\subsection{Holonomous gluon self energy at one loop order}\label{sec:gluon_self}

The result of Eq. (\ref{quadratic_redux2}) is a contribution to the gluon self energy for $a \neq b$,
\begin{equation}
  \Pi_{{\rm cons};\; 00}^{ab,cd}(p^{ab}) = - \; \delta^{ad} \delta^{bc} \; \frac{1}{p_0^{a b}} \;
  \frac{4 \pi }{3} \, g^2 T^3 \, \sum_{e=1}^N  \left( B_3\left(\frac{\theta_a - \theta_e}{2\pi}\right)
      + B_3\left(\frac{\theta_e - \theta_b}{2 \pi}\right) \right)
   \;   ;
   \label{source_gluon_self_energyA}
 \end{equation}
$p_\mu^{a b}=(p_0^{a b},\vec{p})$, Eq. (\ref{define_p0}).
This term  is constant in the spatial momentum $\vec{p}$, and so a $\delta$-function in space.  With a constrained
functional integral, this term only arises in recognizing that $\epsilon^{\rm cl} \neq 0$; with a source,
that the value of the source must be included.
They  arise  in the insertion method just by doing only the Wick contractions that produce a volume term
\cite{KorthalsAltes:1993ca,Dumitru:2013xna,Guo:2014zra,Guo:2018scp,altes_nishimura}.
Then only gluons radiated from the Polyakov loop stay uncontracted, as in Eq. (\ref{quadratic_wilson_lineA})
that leads to Eq. (\ref{source_gluon_self_energyA}).

The gluon self energy satisfies
\begin{equation}
  p_0^{a b} \Pi_{{\rm cons}; \; 00}^{ab,cd}(p^{ab}) = - \; \delta^{ad} \delta^{bc} \;
  \frac{4 \pi }{3} \, g^2 T^3 \, \sum_{a,b,c=1}^N  \left(
    B_3\left(\frac{\theta_a - \theta_c}{2\pi}\right)  +
    B_3\left(\frac{\theta_c - \theta_b}{2 \pi}\right) \right)
   \;  .
   \label{source_gluon_self_energyB}
\end{equation}
The holonomous self energy has been computed to one loop order in perturbation theory
in $\xi=1$ gauge.  Then, unlike for $\xi=1$  in
zero holonomy, the result is not transverse in the external momentum:
\begin{equation}
  p_\mu^{a b} \; \Pi_{{\rm pert}; \; \mu \nu}^{ab,cd}(p^{ab}) = + \; \delta^{\nu 0} \; \delta^{ad} \delta^{bc} \;
   \frac{4 \pi }{3} \, g^2 T^3 \, \sum_{a,b,c=1}^N  \left( B_3(q_a - q_c)  + B_3(q_c - q_b) \right)
   \;.
   \label{one_loop_gluon_self_energy}
\end{equation}

A transverse but non-local self energy is the sum of the non-local source term from
Eq. (\ref{source_gluon_self_energyA}) and the usual local perturbative diagrams.  Clearly
the contributions of Eqs. (\ref{source_gluon_self_energyB}) and (\ref{one_loop_gluon_self_energy})
cancel identically, so that the sum is transverse,
\begin{equation}
  p_\mu^{a b} \; \Pi_{{\rm total}; \ \mu \nu}^{ab,cd}(p^{ab})
  = p_\mu^{a b} \; \left( \Pi_{{\rm pert}; \; \mu \nu}^{ab,cd}(p^{ab}) +
    \Pi_{{\rm cons}; \; \mu \nu}^{ab,cd}(p^{ab}) \right) = 0 \; ,
  \label{transverse_self_energy}
\end{equation}
where $\Pi_{{\rm cons}; \; \mu \nu} = \delta^{\mu 0}\delta^{\nu 0} \Pi_{{\rm cons}; \; 00}$.
This remains valid when $\xi \neq 1$ \cite{altes_nishimura}.

\subsection{Constraints with double traces}
\label{double_trace_constraint}
In this section we show that the same results hold when the constraint is
an arbitrary function of double traces,
\begin{equation}
{\cal S}_{\rm cons} =   i \epsilon \left( B(\Theta) -
    \int \frac{d^3 x}{V}  \sum_{r=1}^\infty c_r \; \left|{\rm tr} \; {\bf L}^r(x) \right|^2
    \right) \; ,
\end{equation}
where $B(\Theta)$ is manifestly $Z(N)$ invariant.  Consequently, if we choose one $\acl$ to satisfy
the constraint, there will be $N$ equivalent vacua which also satisfy the
constraint.  This doesn't preclude us from introducing such a constraint; we do so because in
constructing effective theories, it is natural to use terms which are $Z(N)$ invariant.

Adding this to the action, instead of Eq. (\ref{constraint_eom}) the equation of motion is
\begin{equation}
  \left( \frac{- i \epsilon^{\rm cl}}{V} \right) \;   \sum_{r=1}^\infty \sum_{b=1}^N  i \, c_r \; r \;
  \left( {\rm e}^{ i r (\theta_a - \theta_b)} - {\rm e}^{- i r (\theta_a - \theta_b)} \right)
  = \frac{8 \pi T^3}{3} \sum_{b=1}^N B_3\left(\frac{\theta_a - \theta_b}{2 \pi}\right) \; .
  \label{constraint_eom_double}
\end{equation}

In addition to Eq. (\ref{quadratic_r}), we also need
\begin{equation}
  {\rm tr}\, \left( {\bf L}_2^r(x)\right)_{\rm off}^\dagger =  + \, \frac{g^2}{2 i } \, \sum_{a \neq b = 1}^N
r \, \left( {\rm e}^{- i r \theta_a} - {\rm e}^{- i r \theta_b} \right) \, 
T \sum_{n=-\infty}^{+\infty}\,   \frac{1}{p_0^{a b}}
  \left( \aqu^{ba}(x,p_0) \; \aqu^{ab}(x,-p_0) \right) \; .
  \label{quadratic_r_conj}
\end{equation}
thus at quadratic order the contribution of off-diagonal elements to the Lagrangian is
\begin{eqnarray}
  - \; \left( \frac{- i \epsilon^{\rm cl}}{V} \right) \;
  \frac{g^2}{2 i } \, \sum_{r=1}^\infty c_r \; r \sum_{a \neq b = 1,c=1}^N
  &&\, \left( {\rm e}^{i r (\theta_a - \theta_c)} - {\rm e}^{- i r (\theta_a - \theta_c)} - {\rm e}^{ i r (\theta_b-\theta_c)}
  + {\rm e}^{- i r  (\theta_c - \theta_c)} \right) \, \nonumber \\ 
&& \times \; T  \sum_{n=-\infty}^{+\infty}\,   \frac{1}{p_0^{a b}}
  \left( \aqu^{ba}(x,p_0) \; \aqu^{ab}(x,-p_0) \right) \; .
  \label{quadratic_double}
\end{eqnarray}
By using the equation of motion in Eq. (\ref{constraint_eom_double}), though,
this reduces identically to the result of Eq. (\ref{quadratic_redux2}).  Thus
all of the results obtained previously by constraining terms linear in Polykov loops go through unchanged.
This includes the identity of the free energy to $\sim g^2$ and the transversity of the gluon self
energy.  

As seen previously for a constraint involves linear powers of the Polyakov loop, which was independent
of the ${\cal J}_r$, for constraints with double traces, the gluon self energy is independent of
the specific coefficients that enter, the $c_r$.

Polyakov loops from constraints (or sources)
also contribute to correlation functions of $\aqu$ to higher order.
For example, cubic terms will involve two and three factors of $1/p_0^{a b}$; assuming that the
later cancel, as for the quadratic terms, the same reduction by the equations of motion appears
plausible.  It is natural
to suppose that these will cancel other terms which arise from purely perturbative computations
to $\sim g^3$, {\it etc.}, but we have not explicitly verified this.

We also suggest that similar properties hold for arbitrary functions of Polyakov loops, but the
above suffices for our purposes herein.  Indeed, the generality of these results hints that a more
general property of path ordered loops is at work, which is at present obscure to us.

\section{Free energy to $\sim g^3$}\label{sec:cubic}

In describing the transition to a confined phase, for $\aqu \sim \Theta \, T/g$
we take $\theta_{a} \sim 1$ for all $a$.
Doing so, it is obvious that for the off-diagonal modes, the background field cuts off any
possible infrared divergence  (For quantities like the surface tension,
some off-diagonal $\theta_{ab }$ are {\it per se} vanishing and contribute infra red divergences
\cite{Giovannangeli:2002uv,*Giovannangeli:2004sg,altes2007}.)
In computing the free energy, this is for the static modes, with
$p_0 = 0$.  Thus a ``hard'' field, with $\Theta \sim 1$, the free energy can be expanded in a power
series in $g^2$.

In perturbation theory, it is well known that the static modes are infrared divergent, and
contribute to the free energy at $\sim g^3$.  We consider how a ``soft'' background field, with
$\theta_a \sim g$, contributes to the free energy.  This is thus how the transition holonomous plasma
first emerges from the strict perturbative limit.

If the total self energy at $\sim g^2$ is $\delta \Pi_{\mu \nu}$, then by resumming the
ring diagrams, they contribute to the free energy
\begin{equation}
  {\cal F}_3 = - \; \sum_{a,b=1}^N \; T \sum_{n=-\infty}^{+\infty} \int \frac{d^3p}{(2 \pi)^3}
  \; {\rm tr} \; \log \left( (p^{a b})^2 \delta^{\mu \nu} + (\xi^{-1} - 1) p_\mu^{a b} p_\nu^{a b}
    - \delta \Pi_{\mu \nu}^{a b} \right) \; .
  \label{ring_diagrams}
\end{equation}
To obtain ${\cal V}_3={\cal F}_3$ we take the static mode and drop the contribution of $\delta \Pi$ to linear
order, which is part of the free energy $\sim g^2$.  We simply note that the $\xi$-dependence is
proportional to
\begin{equation}
  \sim \xi^{-1} p_\mu^{a b} p_\nu^{a b} \delta \Pi_{\mu \nu}^{a b} \; .
\end{equation}
From Eq. (\ref{transverse_self_energy}), this vanishes.  

In the perturbative vacuum, the computation of ${\cal F}_3$ is then straightforward.  We take
Feynman gauge, $\xi = 1$, for simplicity.  The most infrared divergent term is clearly from
the static mode, with $p_0 = 0$.  In this limit, the only component of $\Pi^{\mu \nu}$ which is
nonzero is
\begin{equation}
  \Pi^{0 0}(p_0 = 0, \vec{p} \rightarrow 0) = m^2_{\rm Debye} = \frac{g^2 N}{3} \; T^2 \; ,
  \label{debye_mass_zero_holonomy}
\end{equation}
where $m^2_{\rm Debye}$ is the Debye mass, squared.  Integrating over $\vec{p}$,
\begin{equation}
  {\cal F}_3 = - \; \sum_{a,b=1}^N \; T \int \frac{d^3p}{(2 \pi)^3}
  \; {\rm tr} \; \log \left( \vec{p}^2 + m^2_{\rm Debye} \right)
  \sim T \; m_{\rm Debye}^3 \sim g^3 T^4 \; .
  \label{free_energy_3_zero_holonomy}
\end{equation}
Away from $\theta_a = 0$, the results for ${\cal F}_3$ are less obvious.

\subsection{Off-diagonal gluons}
\label{off_diagonal_gluons}

The computation of the self energy to one loop order is given in Ref. \cite{KorthalsAltes:2019yih}.
Here we give an alternate derivation, using 
results from the Hard Thermal Loop (HTL) limit.  Typically, the HTL is
computed after analytically continuing the Eucldiean energy $p_0 \rightarrow - i \omega$;
it is valid for soft momenta, taking both $\omega$ and $|\vec{p}|$ soft, $\sim g T$.

In the Euclidean theory, for the colored momenta $p_0^{a b} = p_0 + T(\theta_a - \theta_b)$ to be
soft requires that $p_0 = 0$ and that all $\theta_a \sim g$.  In the HTL limit, the gluon self energy is
\cite{Hidaka:2009hs}
\begin{equation}
\Pi^{ab,cd}_{{\rm pert}; \mu \nu}(p^{a b}) \approx
-{\cal K}_{\rm pert}^{a b, cd}(\Theta)\; \delta \Gamma^{\mu \nu}(p^{a b})
- \left(m_{\rm pert}^2\right)^{ab,cd}\!(\Theta) \; \delta \Pi^{\mu \nu}(p^{a b}) \; .
\label{gluon_selfG}
\end{equation}
This result is independent both of the gauge fixing parameter, and of the particular gauge
chosen.  The only requirement is that the external momenta are all soft.

The first term involves the function ${\cal K}$, which depends only upon the $\theta_a$'s:
\begin{align}
& \notag {\cal K}^{a b,cd}_{\rm pert}(\Theta) \\ &
= \frac{4  \pi i }{3} g^2 T^3
\left( \delta^{ad} \delta^{b c} \; \sum_{e = 1}^{N} \; 
  \left( B_3\left(\frac{\theta_a - \theta_e}{2\pi}\right)
    + B_3\left(\frac{\theta_e - \theta_b}{2 \pi}\right) \right)
  - 2 \; \delta^{ab} \delta^{c d} B_3\left(\frac{\theta_a - \theta_c}{2 \pi}\right) \;
\right) \; ;
\label{gluon_selfI}
\end{align}
in Ref. \cite{Hidaka:2009hs}, ${\cal A}_0(T \theta) = 2 B_3(\theta/(2 \pi))$ was used.  (This corrects Eq. (158)
of Ref. \cite{Hidaka:2009hs}, where the coefficient on the right hand side should be $2 \pi i g^2 T^3/3$ instead
of $2 g^2 T^3$.)  The soft momenta enter through the function
\begin{equation}
  \delta \Gamma^{\mu \nu}(p^{a b}) = - \; \frac{1}{i \, p_0^{a b}} \; \delta \Pi^{\mu \nu}(p^{a b})
  - \; u^\mu u^\nu \; \frac{1}{i \, p_0^{a b}} \; .
  \label{delta_gamma}
\end{equation}
$\delta \Pi^{\mu \nu}(p)$ is the standard function which appears in Hard Thermal Loops,
\begin{equation}
  \delta \Pi^{\mu \nu}(p) =
\; - u^\mu u^\nu - i p_0\int \frac{d \Omega}{4 \pi} \; \frac{ \hat{K}^\mu \hat{K}^\nu }
{p \cdot \hat{K}}   \; .
\label{landauQ_T}
\end{equation}
The integral is over all directions of the unit spatial vector $\hat{k}$; 
$\hat{K}=(-i,\hat{k})$ is a null vector, $\hat{K}^2 = 0$.  This function remains valid
if $p_0 \rightarrow p_0^{a b} = T (\theta_a - \theta_b)$.

The Debye mass squared for $\theta_a \neq 0$ also enters,
\begin{align}
&\left(m_{\rm pert}^2\right)^{ab,cd}(\Theta) 
\notag \\ & =  g^2 T^2
\left( 
\delta^{ad} \delta^{b c}  \; \sum_{e = 1}^{N} \; 
\left( B_2\left(\frac{\theta_a - \theta_e}{2 \pi}\right)
  + B_2\left(\frac{\theta_e - \theta_b}{2 \pi}\right) \right)
 - 2\; \delta^{ab} \delta^{c d} \; B_2\left(\frac{\theta_a - \theta_c}{2 \pi}\right) 
\right)
\; ;
\label{gluon_selfH}
\end{align}
in \cite{Hidaka:2009hs}, ${\cal A}(T \theta) = 6 B_2(\theta/(2 \pi))$ was used.

After analytic continuation, the above expressions apply for soft $\omega$ and $p$,
and arbitrary $\theta_a \sim 1$.  To compute
${\cal F}_3$, we need the limit in which $p_0 = 0$ and all $\theta_a \sim g$.  In this limit, we can approximate
$B_2(0) = 1/6$, and $B_3(x) \approx x/2$.
In $\delta \Gamma^{\mu \nu}$, the term $2 \pi i B_3((\theta_a-\theta_b)/(2\pi))/ip_0^{a b} \sim 1/2$ at small $\theta_a$.
Doing so, we find that {\it all} terms
$\sim \delta \Pi^{\mu \nu}(p^{a b})$ cancel {\it identically}.  This only leaves the
the term $\sim - u^\mu u^\nu/(i p_0^{a b})$ on the right hand side of Eq. (\ref{delta_gamma}).
However, this enters proportional to $B_3((\theta_a - \theta_e)/(2\pi)) + B_3((\theta_e - \theta_b)/(2\pi))$.
By the previous analysis
in Eq. (\ref{source_gluon_self_energyA}), this also cancels against the contribution
of the constraint term, $\Pi_{{\rm cons}; \; \mu \nu}$.  

This implies that for small $\Theta$, {\it all} contributions to the self energy for off-diagonal gluons
vanish for $p_0 = 0$ and $\Theta \sim g$.  This cancellation only occurs for small $\Theta$, and does
not hold when $\Theta \sim 1$.  

This does not imply that there are long ranged fields.  
In the presence of the background $\acl$ field,
at leading order the inverse propagator for the transverse gluons is
\begin{equation}
  \Delta^{-1} = (p_0 + T(\theta_a - \theta_b) )^2 + \vec{p}^2 \; .
\end{equation}
Thus even for static modes with $p_0 = 0$, a nonzero holonomy, $\theta_a \neq 0$, acts like a mass term.

The analysis implies that static electric fields are not screened for small $\Theta$.  This can also be
seen from the transversity of the total gluon self energy, $\Pi_{{\rm total}; \; \mu \nu}^{ab,cd}(p^{ab})$
in Eq. (\ref{transverse_self_energy}).  In the static limit, as $\vec{p} \rightarrow 0$ this reduces
to
\begin{equation}
  (\theta_a - \theta_b) \; \Pi_{{\rm total}; \; 00}^{ab,cd}(2 \pi T(\theta_a - \theta_b),0) = 0 \; .
\end{equation}
Consequently, when $\theta_a - \theta_b \neq 0$, the self energy vanishes, 
$\Pi_{{\rm total}\; ; 00}^{a b,cd}(T(\theta_a - \theta_b),0) = 0$.
Clearly, for this to hold, it is essential that the gluon self energy is transverse.

This is very different from when $\theta_a = 0$; then going to the static
limit does not constrain $\Pi_{00} \sim m_{\rm Debye}^2$.  From Eq. (\ref{debye_mass_zero_holonomy}),
$m_{\rm Debye}^2 \neq 0$, so static electric fields are screened when $\theta_a = 0$.

This behavior can be derived directly without explicit evaluation of the one loop diagrams.
We use the expressions for the Hard Thermal Loops in a holonomous plasma \cite{Hidaka:2009hs}.
This only applies for soft momenta, so both the spatial momentum $p$ and the $\theta_a$ are soft, $\sim g$.  
Consider the diagram with two three gluon vertices.  
After summing over the loop momentum $k_0$, from Eqs. (115) and (116) of Ref. \cite{Hidaka:2009hs}
the contribution to $\Pi^{i j}$ is proportional to
\begin{equation}
{\cal J}^{i j}(p, T \theta_1, T \theta_2) \sim
\; \int^\infty_0 d^3 k \; \frac{k^i k^j}{E_k E_{p - k}} \;
 \int \frac{d \Omega}{4 \pi} \; 
\left( {\cal I}_2 + {\cal I}_3 \right) \; .
\label{landauQ_B1}
\end{equation}
We assume that the loop momentum $k$ is hard, so in each three gluon vertex we can take $\sim k^i$, dropping
terms $\sim p^i$.  Similarly, we approximate $E_{p-k} \sim k$.
Then the momentum dependence arises entirely from the statistical distribution 
functions and from the energy denominators.  This is given by
\begin{equation}
{\cal I}_2 =
\frac{n(E_k - i T \theta_1) - n(E_{p - k} + i T \theta_2) }
{ip_0^{1 2} - E_k + E_{p - k}} \;\;\; ; \;\;\;
  {\cal I}_3  = \; \frac{n(E_{p-k} - i T \theta_2) - n(E_{k} + i T \theta_1) }
{ip_0^{1 2} + E_k - E_{p - k}} \; ,
\label{landauQ_C}
\end{equation}
where $p_0^{1 2} = p_0 + T (\theta_1 +  \theta_2)$.

These factors arise from Landau damping, and involve a difference of hard energies.  The difference is
a soft energy, so we need to expand $E_{p-k} \approx k - \hat{k} \cdot \vec{p} + \ldots$.
Since we are computing the self energy for Euclidean momentum, we work 
in the static limit, $p_0 = 0$.  For simplicity we assume $\theta_1 = 0$ and $\theta_2 = \theta$.  Under
these approximations,
\begin{equation}
  {\cal I}_2 \approx \frac{1}{- \hat{k} \cdot \vec{p} + i T \theta} \left(
     n(k) - n(k - \hat{k} \cdot \vec{p} + i T \theta) \right) \approx - \frac{d}{d k} n(k) \; .
\end{equation}
and
\begin{equation}
  {\cal I}_3 \approx  \frac{1}{ \hat{k} \cdot \vec{p} + i T \theta} \left(
    n(k - \hat{k} \cdot \vec{p} - i T \theta) - n(k) \right) \approx - \frac{d}{d k} n(k) \; .
\end{equation}
Each term is nonzero, but the point is that it is independent of {\it both} the external spatial momentum,
$\vec{p}$, and the holonomy, $T \theta$.  This is most unexpected, as it is certainly possible for the result
to depend upon the dimensionless ratio $|\vec{p}|/(\theta T)$.  

It is also useful to consider the behavior of the free energy.
At one loop order any mode with nonzero energy, $p_0 \neq 0$, clearly
contributions to the determinant,
${\rm tr} \log \Delta^{-1}$, are regular about $\theta_a = 0$.  Thus modes
with nonzero energy contribute
only to the terms quadratic and quartic in the $\theta_a$'s:, $\sim (\theta_a - \theta_b)^2$ and
$\sim ((\theta_a - \theta_b)^2)^2$ in $B_4((\theta_a - \theta_b)/(2\pi))$, Eqs. (\ref{potential_one_loop}) and
(\ref{def_B4}).  There is also a cubic term in  $B_4((\theta_a - \theta_b)/(2\pi))$, $\sim ((\theta_a - \theta_b)^2)^{3/2}$;
it is easy to show that this arises uniquely from the mode with static energy, $p_0 = 0$.
Thus the origin of the cubic term in the one loop potential
is similar as that of ${\cal F}_3$ when $\theta_a = 0$, Eq. (\ref{free_energy_3_zero_holonomy}).
When the $\theta_a$ are soft, this cubic term at one loop order is $\sim g^3$, like that in perturbation
theory.  Similarly, those at two loop order are $\sim g^5$.  What is unexpected is that the
free energy does not appear to be continuous as $\Theta \rightarrow 0$: there are cubic terms $\sim |\theta_a|^3$
when $\theta_a \neq 0$, but these vanish as $\theta_a \rightarrow 0$.  In contrast, at zero holonomy
there {\it is} a cubic term $\sim g^3$.  


\subsection{Diagonal elements to $\sim g^3$}\label{sec:diagonal}

\label{diagonal_gluons_g3}

In principle, the computation of the contribution of color diagonal gluons to the free energy
at weak holonomy is straightforward.  As argued previously, gauge invariant sources must be
used, minimized with respect to the background field, and the Debye masses in the presence of
the background field computed.

We show that when the explicit potentials are computed, that a surprise arises.  Because
the potential at one loop order involves a sum over an infinite number of loops, any source
must also involve an infinite sum, of a specific form.

Our arguments can be made precise for two and an infinite number of colors.  After treating
these two examples in detail, we discuss arbitrary $N$.  

\subsubsection{Two colors}
\label{sec:two}

For two colors, define $\theta_1 = - \theta_2 = \pi \, q$.  To one loop order the perturbative potential is
\begin{equation}
  {\cal V}_1(q) = \pi^2 T^4 \left( - \frac{1}{15} + \frac{4}{3} \; q^2 (1-q)^2 \right) \; .
  \label{pert_two_colors}
\end{equation}
We set $T=1$ for convenience.  Normalized to unity, the Polyakov loop $\ell = \cos(\pi q)$, with
$q=0$ the perturbative vacuum, and $q= 1/2$ the confined.  We add two sources,
\begin{equation}
  {\cal V}_{j}(q) = 4 \, j_1 \, (\ell^2 -1) + 16 \, j_2 \,(\ell^4 -1)\; .
  \label{nonpert_two_colors_loops}
\end{equation}
The potential with just $j_1$ was considered in Ref. \cite{Dumitru:2012fw};
that with $j_1$ and $j_2$ was discussed in Ref. \cite{Nishimura:2011md}.
The total potential is then
\begin{equation}
  {\cal V}_{\rm tot}(q) = {\cal V}_1(q) + {\cal V}_{j}(q) \; .
\end{equation}
For large values of $j_1$ and $j_2$ the potential minimizes
the loop, and drives the theory to the confined vacuum, $q = 1/2$.
Our interest is how this occurs.

Begin with $j_2 = 0$.  As $j_1$ increases, there is a transition from
$q=0$ to $q \neq 0$ at
\begin{equation}
  j_1^0 = \frac{\pi^2}{48} \; , \; j_2 = 0 \; .
\end{equation}
This transition is of first order, directly to the confining vacuum with $q = 1/2$
\cite{Dumitru:2012fw, Nishimura:2011md}.

This is {\it not} what we require, however, but rather a transition to a nonzero but arbitrarily small value of $q \neq 0$.
Consider expanding about the confined phase, with $q = 1/2$:
\begin{equation}
  {\cal V}_{\rm tot}\left(\frac{1}{2} - \delta q \right)  \approx 
  \frac{\pi^2}{12} - 4\, j_1 - 16 \, j_2  + 4 \, \pi^2 \left(j_1 - \frac{1}{6} \right) \delta \theta^2 
             + \frac{4}{3} \pi^4 \left( - \frac{1}{\pi^2} + j_1 - 12 j_2 \right) \delta \theta^4 + \ldots
  \label{pot_two_conf}
\end{equation}
Thus there is a line of second order transitions from the deconfined to the confined phase
when $j_1 = 1/6$.

For the quartic coupling to be positive \cite{Nishimura:2011md},
\begin{equation}
  j_2 \geq \frac{1}{12} \left( \frac{1}{6} - \frac{1}{\pi^2} \right) \; .
\end{equation}
This is not sufficient: at $j_1 = 1/6$ and this value of $j_2$,
the value of the potential at $q = 1/2$ is higher than at $q = 0$, not lower.

\begin{figure}
  \includegraphics[width=0.6\linewidth]{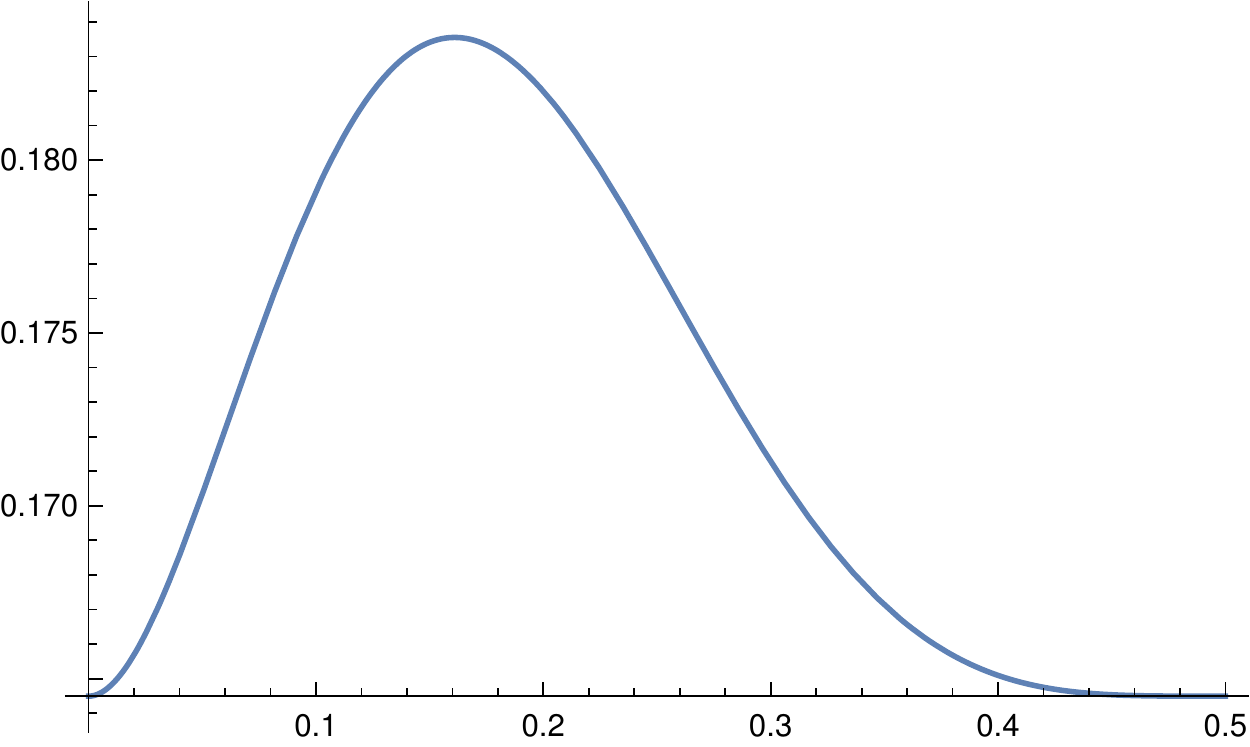}
  \caption{A plot of the potential for the critical first order
    point, Eq. (\ref{critical_first}).  The
    masses squared is nonzero about the perturbative vacuum, $q=0$, but vanishes about
    the confining vacuum, $q=1/2$.  
  }
  \label{fig:critical_first}
\end{figure}

Fix $j_1 = 1/6$, and move up in $j_2$ to
\begin{equation}
  j_1^{\rm crit} = \frac{1}{6}    \; , \; j_2^{\rm crit} = \frac{1}{16}
  \left( \frac{\pi^2}{12} - \frac{2}{3} \right) \; .
  \label{critical_first}
\end{equation}
At this point, the potential has an uncommon form, illustrated
in Fig. (\ref{fig:critical_first}).  The value of the potential
is equal at $q= 0$ and $q = 1/2$, with a barrier between them, and so the transition is of first order.
Nevertheless, the mass in the confining vacuum,
at $q = 1/2$, vanishes.  We call this a critical first order transition;
they also occur at infinite $N$ in
some matrix models \cite{Dumitru:2004gd,Pisarski:2012bj}.  It occurs for two colors because the potential is not
just a simple polynomial in $q$.

Moving up in $j_2$ for a constant value of $j_1 = 1/6$,
there is a standard second order transition.
Now consider first increasing $j_1$ from $j_1^{crit}$.
From Eq. (\ref{pot_two_conf}), the potential at $q= 1/2$ vanishes along
the straight line
\begin{equation}
  j_2 = \frac{1}{4} \left( j_1^0 - j_1\right) \; .
\end{equation}
Along this line, there is a first order transition from $q = 0$ directly to the confining vacuum,
$q = 1/2$.

The behavior for $j_1 < j_1^{\rm crit}$ is more involved.  In this case we take
\begin{equation}
  j_1 = j_1^{\rm crit} - \delta j_1 \;\;\; , \;\;\; j_2 = \frac{1}{4} \, \delta j_1 + \delta j_2 \; .
\end{equation}
Expanding about this point, we find a first order transition from $q = 0$ to $q = 1/2 - \delta q$,
\begin{equation}
  (\delta q)^2 = 8 \, a \, \delta j_1 \;\;\; , \;\;\;
  \delta j_2 = - \, a \, \delta j_1^2 \;\;\; , \;\;\; a = \frac{9 \pi^2}{3 \pi^4 + 16} \; .
\end{equation}
That $\delta q \sim \sqrt{\delta j_1}$, instead of $\delta q \sim \delta j_1$, follows because
the mass vanishes about the confined phase at $j_1^{\rm crit}$.  

Thus there is a line of first order transitions as $j_1$ decreases.
Along this line, there is a first order transition from $q = 0$ to a value of $q_0 < 1/2$.
This line goes down to
$j_1 = 0$, where there is a first order transition at $j_1 = 0$ and $j_2^0 \approx 0.03615\ldots$;
at this point, the minimum of the potential jumps from $q = 0$ to $q_0 \approx 0.145\ldots$.

\begin{figure}
  \includegraphics[width=0.8\linewidth]{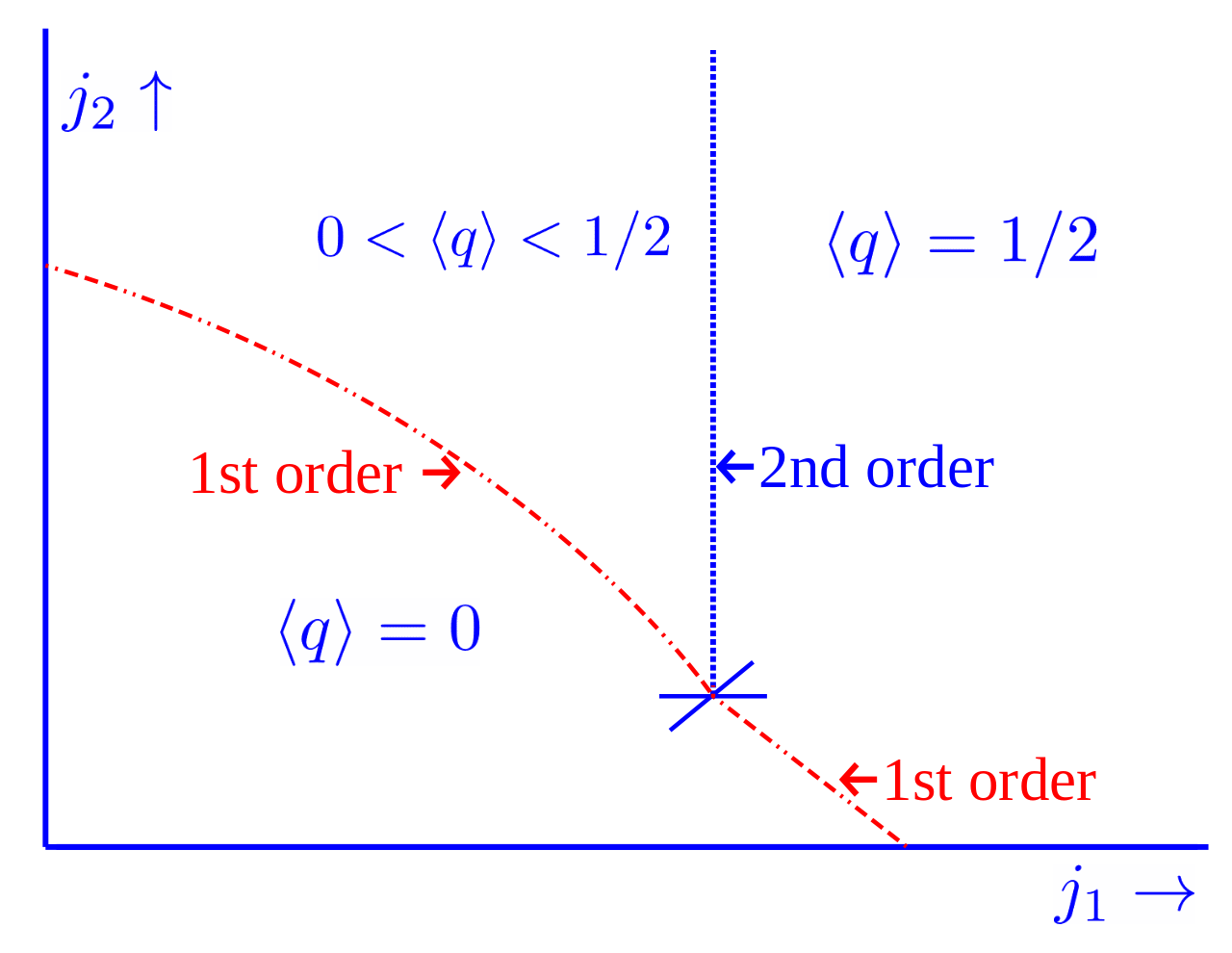}
  \caption{The phase diagram for the effective model for two colors, with the potential
    of Eq. (\ref{pot_two_conf}).  There are three regions: strict perturbative, with $\langle q_a \rangle = 0$;
    holonomous plasma, with $0 < \langle q_a \rangle < 1/2$, and confined, with $\langle q_a \rangle = 1/2$.
    The cross denotes $(j_1^{\rm crit},j_2^{\rm crit})$, Eq. (\ref{critical_first}).  Note that there is an
    unbroken line of first order transitions betweeen the confined and perturbative phases.
  }
  \label{fig:two_colors}
\end{figure}

This gives rise to the phase diagram of Fig. (\ref{fig:two_colors}).
There is an {\it un}broken line of first order transitions, with {\it no} smooth transition
from $\langle q \rangle = 0$ to a nonzero value.  
This phase diagram is qualitatively different form Fig. (1) in Ref. \cite{Nishimura:2011md},
where the line of first order transitions terminates at $j_1^{\rm crit}$.

Thus the two sources used in Eq. (\ref{nonpert_two_colors_loops}) are not adequate to generate
a small value of $\Theta$ for arbitrarily small sources.  Consider the expansion for small $q$.
Then the sources, as functions of $\cos(\pi q)$, begin at quadratic order.  The same is true
for the potential at one loop order, but in addition, there is a term of cubic order, with a negative
sign.  The terms from the sources can be tuned so that the coefficient of the quadratic term vanishes,
but that still leaves negative cubic term, which drives a first order transition.

This argument is unavoidable for two colors, and can be immediately generalized to three colors.  We comment
that the appearance of a cubic term, which implies non-analyticity in $q$, is because the potential involves
a sum over an infinite number of loops.

\subsubsection{Infinite colors}
\label{sec:infinite}

For four or more colors, there is more than one independent $\theta_a$.  
While the presence of a cubic
term in the perturbative potential, $B_4(\theta_a -\theta_b)$, suggests that one cannot smoothly move from
$\theta_a = 0$ to nonzero $\theta_a$, it is not evident that it might not happen for one special
direction of the $\theta_a$.  

In this subsection we compute for an infinite number of colors, using
standard techniques for matrix models at large $N$
\cite{Brezin:1977sv, Gross:1980he, *Wadia:1980cp, Lang:1980ws, *Menotti:1981ry, Jurkiewicz:1982iz,
Green:1983sd, Damgaard:1986mx, Azakov:1986pn,Demeterfi:1990gb,*Jurkiewicz:1990we, Sundborg:1999ue,Aharony:2003sx, *Aharony:2005bq, Dumitru:2003hp, Dumitru:2004gd, AlvarezGaume:2005fv, Schnitzer:2004qt, Hollowood:2009sy, *Hands:2010zp, *Hollowood:2011ep, *Hollowood:2012nr, Ogilvie:2012is, Liu:2015yaa},
and in particular
using the known solution for this particular model \cite{Pisarski:2012bj,Nishimura:2017crr}.
We revert to using the $\theta_a$, as in Refs. \cite{Pisarski:2012bj,Nishimura:2017crr} 
At large $N$ we replace the discrete label $a$ by a continuous index $x$,
where $x = a/N - 1/2$, and introduce the eigenvalue density,
\begin{equation}
  \rho(\theta) = \frac{dx}{d\theta} \; .
\end{equation}
At large $N$ the one loop potential is $N^2$ times
\begin{equation}
  {\cal V}_1(\theta) \sim \int^\pi_{-\pi} d\theta_1 \int^\pi_{-\pi}  d \theta_2
  \; \rho(\theta_1) \; \rho(\theta_2) \;
  |\theta_1 - \theta_2|^2 \left(1 - \frac{|\theta_1 - \theta_2|}{2 \pi}\right)^2 \; ,
\end{equation}
up to an overall constant, Eq. (\ref{potential_one_loop}).  
For small $\theta$ we certainly expect a cubic term of negative sign, but to establish this definitively
requires the explicit solution for the eigenvalue density \cite{Nishimura:2017crr}.
Previous study concentrated on the transition from
the confined to the deconfined phase, but the analysis can be adapted to
how the theory leaves the perturbative limit.

In terms of the eigenvalue density the $n^{\rm th}$ Polyakov loop equals
\begin{equation}
  \ell_n = \frac{1}{N} \; {\rm tr} \; {\bf L}^n
  = \int^{\pi}_{- \pi} d \theta \; \rho(\theta) {\rm e}^{i n \theta} \; .
\end{equation}
We assume that by an overall $Z(N)$ rotation the expectation value of all loops is 
real, so the eigenvalue density $\rho(\theta)$ is an even function in $\theta$.

The perturbative potential can be rewritten in a power series in Polyakov loops, Eq. (\ref{potential_one_loop}).
Notice that the overall sign is negative, so the potential is minimized when all loops are maximal:
all $\ell_n = 1$, so all $\theta_a(x) = 0$.

In studying the transition from the confined to the deconfined phase, Ref. \cite{Nishimura:2017crr}
assumed that the coefficient of Eq. (\ref{potential_one_loop}) is positive.  The
potential is then minimized when all loops vanish, which is the confined phase.

The eigenvalue density for large $N$ is soluble regardless of the overall sign of Eq. (\ref{potential_one_loop}).
In the notation of Ref. \cite{Nishimura:2017crr}, the solution is the case $s = 4$.
While the effective potential is a function of all $\ell_n$, we can integrate out all loops except for the
first, $\ell_1$.  It is necessary to introduce an external field for $\ell_1$, $\omega$.  The potential
for $\omega$ is
\begin{equation}
  {\cal F}_1(\omega) = + \; 2 \; \int^\omega_0 \, d \omega' \; \ell_1(\omega') \; .
  \label{def_v1_omega}
\end{equation}
The sign on the right hand side is positive, opposite to the negative sign in
Eq. (31) of Ref. \cite{Nishimura:2017crr}.  This is due to 
overall change in sign of the potential in Eq. (\ref{potential_one_loop}).

The solution for the eigenvalue density is \cite{Nishimura:2017crr}
\begin{equation}
\rho(\theta) =
\frac{1}{2 \pi}
\left(
  \left(\pi - \theta_0 - 2 \omega \sin \theta_0 \right)
  \left( \delta(\theta-\theta_0)+\delta(\theta+\theta_0) \right)
  +1+ 2\omega \cos \theta
\right) 
  \, 
\label{rho_s4}
\end{equation}
The is defined for $|\theta| \leq \theta_0$, with two $\delta$-function singularities at each end, for
$\theta = \pm \theta_0$.  The solution
vanishes when $|\theta| > \theta_0$, which we call a single gap (as $\rho(\theta)$ is even in $\theta$).
The eigenvalue density can be computed for $s = 1$, $2$, $3$, and $4$, but the singularities
at $\theta = \pm \theta_0$ are special to $s=4$ \cite{Nishimura:2017crr}.

The endpoint of the gap, $\theta_0$, is related to the background field, $\omega$, through the relation
\begin{equation}
  \omega  = \frac{1}{6}
  \left(
    \frac{\left(\pi-\theta_0\right)^3 }{ \sin \theta_0+ \left(\pi-\theta_0\right) \cos \theta_0}
    \right) \; .
\label{boundary_p4}   
\end{equation}
The strict perturbative limit is when $\theta_0 = 0$, as
\begin{equation}
\rho_{\rm pert}(\theta) = \delta(\theta) \;\;\; ; \;\;\;   \omega = \omega_c = \frac{\pi^2}{6} \; .
\end{equation}
Expanding in $\omega$,
\begin{equation}
  \omega = \frac{\pi^2}{6} - \delta \omega \; ,
\end{equation}
the solution of Eq. (\ref{boundary_p4}) is
\begin{equation}
  \theta_0(\omega) = \frac{2}{\pi}
  \left( \delta \omega + \frac{1}{3} \left( 1 + \frac{6}{\pi^2} \right)\, \delta \omega^2
    + \frac{2}{9} \left(1 + \frac{1}{\pi^2} + \frac{30}{\pi^4} \right) \delta \omega^3
    + \frac{5}{27} \left( 1 - \frac{1}{\pi^2} + \frac{144}{\pi^6} \right) \delta \omega^4 +
    \ldots \right)
    \; .
\end{equation}
With the eigenvalue density of Eq. (\ref{rho_s4}), the first Polyakov loop equals
\begin{equation}
  \ell_1 (\omega) = \frac{1}{\pi} \left( \omega \, \theta_0+ \sin \theta_0 + \cos \theta_0
    \left(\pi - \theta_0 - \omega \sin \theta_0 \right) \right) \; .
\label{rho1_s4}
\end{equation}
Loops with $n \geq 2$ are given by
\begin{equation}
  \ell_{n \geq 2} =
  \frac{1}{\pi} \left(
    (\pi - \theta_0) \cos(n \theta_0) + \frac{\sin(n \theta_0)}{n}
    + \frac{2 \, \omega \, n}{n^2-1} 
      ( - n \, \sin \theta_0 \, \cos(n \theta_0 ) + \cos\theta_0 \, \sin(n \theta_0 )
    \right)
    \; .
    \label{higher_loop_th}
\end{equation}
We need to compute these loops not as functions of $\theta_0$, but of the external field, $\omega$.
For $\ell_1$, 
\begin{equation}
  \ell_1(\delta \omega) \approx 1 - b_2 \, \delta\omega^2 - b_3 \, \delta\omega^3 - b_4 \, \delta\omega^4 + \ldots \; ,
  \label{rho1_dom}
\end{equation}
where
\begin{equation}
  b_2 = \frac{2}{\pi^2} \;\; , \;\; b_3 = \frac{4}{9 \pi^2} \left(1 + \frac{12}{\pi^2} \right) \;\; , \;\;
  b_4 = \frac{2}{9 \pi^2} \left( 1 + \frac{1}{\pi^2} + \frac{84}{\pi^2} \right) \; .
\end{equation}
Solving for ${\cal F}_1$ from Eqs. (\ref{def_v1_omega}),
\begin{equation}
  {\cal F}_1(\omega) - {\cal F}_1(\omega_c) = 2 \left( \delta \omega - \frac{b_2}{3} \, \delta\omega^3
    - \frac{b_3}{4} \, \delta \omega^4 + \ldots \right) \; .
\end{equation}
The potential, as a function of $\ell_1$, is given by
\begin{equation}
  {\cal V}_1(\ell_1) - {\cal V}_1(1) =
  \left. {\cal F}_1(\omega) + 2 \, \omega \, \ell_1(\omega)\right|_{\omega = \omega(\ell_1)}
  - \left( {\cal F}_1(\omega_c) + 2 \, \omega_c  \right) \; .
\end{equation}
As a function of $\omega$,
\begin{equation}
  {\cal V}_1(\ell_1) - {\cal V}_1(1) 
  \approx \frac{2}{3} \, \delta\omega^2 + \frac{4}{27} \left( 1  - \frac{6}{\pi^2} \right) \delta\omega^3
  + \frac{2}{27} \left(-1 + \frac{8}{\pi^2} + \frac{24}{\pi^4} \right) \delta\omega^4 + \ldots \; .
  \label{potential_fnc_om}
\end{equation}

The right hand side is a function of $\delta \omega$, but it is necessary to invert Eq. (\ref{rho1_dom})
and write it as a function of $\ell_1$.  We introduce
\begin{equation}
  (\delta \rho)^2 = \frac{\pi^2}{2} \, (1 - \ell_1) \; ;
  \label{def_delta_rho}
\end{equation}
this $\delta \rho$ is, by definition, a measure of the deviation from zero for all eigenvalues.
Eq. (\ref{rho1_dom}) gives
\begin{equation}
  \delta \omega = \delta \rho - \frac{1}{9} \left(1 + \frac{12}{\pi^2} \right)\delta \rho^2
  + \frac{2}{81} \left( - 1 + \frac{111}{4 \pi^2} - \frac{9}{\pi^4} \right) \delta \rho^3 + \ldots \; .
  \label{om_func_dlr}
\end{equation}
Substituting this into Eq. (\ref{potential_fnc_om}), we obtain
\begin{equation}
  {\cal V}_1(\ell_1) - {\cal V}_1(1) \approx
  \frac{2}{3} \; \delta (\rho)^2 - \frac{8}{3 \pi^2} \; (\delta \rho)^3
  + \frac{2}{9 \, \pi^2} \left( 1 + \frac{12}{\pi^2} \right) (\delta \rho)^4 + \ldots \; .
\end{equation}
As a function of $\delta \omega$, the $n^{\rm th}$ Polyakov loop of Eq. (\ref{higher_loop_th}) is
\begin{equation}
  \ell_n(\delta \omega) = 1 - \frac{2 }{\pi^2} \, n^2 \, (\delta\omega)^2
  - \frac{4(12 + \pi^2)}{9 \pi^4} \, n^2 \, (\delta\omega)^3
  - \frac{2}{9 \pi^6} (84\, n^2  + (4 - 3 \, n^2) \, n^2 \, \pi^2 + n^2 \, \pi^4) (\delta\omega)^4 + \ldots \; .
  \label{rhon_omega}
  \end{equation}
Using Eq. (\ref{om_func_dlr}),
\begin{equation}
  \ell_n(\ell_1) \approx 1 - \frac{2}{\pi^2} \, n^2 (\delta \rho)^2 + \frac{2}{3 \, \pi^4} \, n^2 (n^2-1) \, (\delta \rho)^4
  + \ldots \; .
  \label{approx_rhon}
  \end{equation}
From Eq. (\ref{def_delta_rho}), $\delta\rho$ is defined as the deviation of $\ell_1$ from unity, and so
all terms of higher order in $\delta \rho$, vanish.  This explains why the quartic term in $\ell_1$ vanishes.
For any $n$, while there is a term cubic in $\delta \omega$ in Eq. (\ref{rhon_omega}),
there is no term cubic in $\delta \rho$.  This accords with the intuition that in expanding about
the perturbative vacuum, that it is an expansion in even powers of the $\theta_a$, and hence of
$\delta \rho$.  

Consider adding a constraint (or source) to a general form for the effective potential of $\ell_1$,
\begin{equation}
  {\cal V}_{\rm eff}(\ell_1) = {\cal V}_1(\ell_1) +   \sum_{n,m=1}^\infty c_n^m \, \ell_n^{2 m} \; .
  \label{general_eff_pot}
\end{equation}
The term quadratic in $\delta \rho$ vanishes when
\begin{equation}
  \sum_{n,m} c_n^m n^2 = \frac{\pi^2}{6} \; .
  \label{critical_condition}
\end{equation}
At this point,
\begin{equation}
  {\cal V}_{\rm eff}(\ell_1) - {\cal V}_{\rm eff}(1) \approx - \, \frac{8}{3 \pi^2} \, (\delta \rho)^3
+ O(\delta\rho^4) \; .
\end{equation}
Thus at the point where the term quadratic in $\delta \rho$ vanishes,
there is a term cubic in $\delta \rho$, with negative sign.
This implies that there is a transition of first order before this point is reached.
The presence of a cubic term is nontrivial, as any single Polyakov loop does not have such
as term, Eq. (\ref{approx_rhon}).  It is the natural extension of the cubic term in the $\theta_a$'s,
expressed in terms of the correct eigenvalue density.

There is a caveat to the above.  In adding terms proportional to the second or higher Polyakov loops,
the eigenvalue densities sometimes develop solutions with two or more
gaps \cite{Jurkiewicz:1982iz}.  We ignore this possibility,
but it seems reasonable to suggest that even such multi-gap solutions will exhibit the first
order transition above.  

Consequently, as for the case with two colors, for any constraint with a finite number of Polyakov loops,
there is a solid region of nonzero measure where the strict perturbative regime holds, with a first
order transition to a holonomous plasma.

\subsubsection{Potentials}

In Sec. (\ref{sec:two}) we showed explicitly that a potential involving the two sources of Eq. (\ref{nonpert_two_colors_loops})
necessarily involves a first order transition.  This is immediately generalized to any finite number of loops,
due to the cubic term in the perturbative potential.  We showed in the
previous section that this remains valid for an infinite number of colors.  It is natural to assume this is true
for any $N$.

What is required is a source which is linear in $\Theta$ for small $\Theta$.  Consider the Bernoulli polynomial,
\begin{equation}
  B_2(\Theta) \sim \sum_{n=1}^\infty \frac{1}{n^2} |{\rm tr} {\bf L}^n|^2 \; .
\end{equation}
In this case, Eq. (\ref{critical_condition}) naively diverges.  The sum is
given in Eq. (\ref{other_bernoullis}), and has a term linear in the $\theta_a$ for small $\theta_a$.

This is true for any $B_n(x)$ when $n$ is odd.  However, the odd $B_n(x)$ are also odd in $x$, and any term
added to the action must be even in $x$.  This suggests that $B_2(x)$ is a natural term to use either
as a source, or as a non-perturbative potential in effective models
\cite{dumitru_how_2011,*dumitru_effective_2012,sasaki_effective_2012,pisarski_gross-witten-wadia_2012, *lin_zero_2013,kashiwa_critical_2012,*kashiwa_roberge-weiss_2013,*gale_production_2015,*hidaka_dilepton_2015,*satow_chiral_2015,*lin_collisional_2014,Pisarski:2016ixt,*Folkestad:2018psc}.
Of course this does not imply that $B_2(x)$ must be used, only that any such potential
must involve a sum over an infinite number of Polyakov loops, and have a term linear in $\theta_a$ about the origin.

\section{Conclusions}

In this paper we have considered the behavior of the free energy at nonzero holonomy in perturbation theory,
and shown that this is not an academic exercise.  Requiring that the source of nontrivial
holonomy is gauge invariant requires that it is a sum over an infinite number of Polyakov loops.
What is more surprising is that the free energy is discontinuous as the holonomy vanishes \cite{KorthalsAltes:2019yih}
to $\sim g^3$.  In a separate work, the BRST identities are used to analyze the free energy
to $\sim g^4$ as the holonomy vanishes \cite{altes_nishimura}.  

This could be merely a peculiar feature of generating non-zero holonomy through an
external source.  It is expected that non-zero holonomy is generated dynamically, as on a femto-torus
\cite{Poppitz:2008hr,*Shifman:2009tp,*Poppitz:2012sw,*Dunne:2016nmc,*Kanazawa:2017mgw}.
Thus it may be that the free energy is continuous as the holonomy vanishes, if it is generated dynamically.
This will be investigated in future work.

\acknowledgments

R.D.P. is funded by the U.S. Department of Energy 
under contract DE-SC0012704. H.N. was supported by the Special Postdoctoral Researchers program of RIKEN and the Japan Society for the Promotion of Science (JSPS) Grant-in-Aid for Scientific Research (KAKENHI) Grant Number 18H01217.
V.S. is funded by the U.S. Department of Energy
under contract DE-SC0020081. 

C.P.K.A. thanks R.D.P. and BNL  for  hospitality and support.

\bibliography{ghosts,qks}

\end{document}